# Solving the Synthetic Riddle of Colloidal 2D PbTe Nanoplatelets with Tunable Near-Infrared Emission

*Leon Biesterfeld [a,b,c], Mattis T. Vochezer [a], Marco Kögel [d], Ivan A. Zaluzhnyy [e], Marina Rosebrock [b,c], Lars F. Klepzig [b,c], Wolfgang Leis [f], Michael Seitz [f], Jannik C. Meyer [d,e], Jannika Lauth [*a,b,c]*

a – Institute of Physical and Theoretical Chemistry, Eberhard Karls University of Tübingen, Auf der Morgenstelle 18, D-72076 Tübingen, Germany.

b – Cluster of Excellence PhoenixD (Photonics, Optics, and Engineering – Innovation Across Disciplines), Welfengarten 1A, D-30167 Hannover, Germany.

c – Institute of Physical Chemistry and Electrochemistry, Leibniz University Hannover, Callinstr. 3A, D-30167 Hannover, Germany.

d – Natural and Medical Sciences Institute at the Eberhard Karls University of Tübingen, Markwiesenstr. 55, D-72770 Reutlingen, Germany.

e – Institute of Applied Physics, Eberhard Karls University of Tübingen, Auf der Morgenstelle 10, D-72076 Tübingen, Germany

f – Institute of Inorganic Chemistry, Eberhard Karls University of Tübingen, Auf der Morgenstelle 18, D-72076 Tübingen, Germany.

ABSTRACT

Near-infrared emitting colloidal two-dimensional (2D) PbX (X = S, Se) nanoplatelets have emerged as interesting materials with strong size quantization in the thickness dimension. They act as model systems for efficient charge carrier multiplication and hold potential as intriguing candidates for fiber-based photonic quantum applications. However, synthetic access to the third family member, 2D PbTe, remains elusive due to a challenging precursor chemistry. Here, we report a direct synthesis for 2D PbTe nanoplatelets (NPLs) with tunable photoluminescence (PL, 910 – 1460 nm (1.36 – 0.85 eV), PLQY 1 - 15 %), based on aminophosphine precursor chemistry. *Ex-situ* transamination of tris(dimethylamino)phosphine telluride with octylamine is confirmed by $^{31}$P NMR and yields a reactive tellurium precursor for the formation of 2D PbTe NPLs at temperatures as low as 0 °C. The PL position of the PbTe NPLs is tunable by controlling the Pb:Te ratio in the reaction.

GIWAXS confirms the 2D geometry of the NPLs and the formation of superlattices. The importance of a post-synthetic passivation of PbTe NPLs by PbI$_2$ to ensure colloidal stability of the otherwise oxygen sensitive samples is supported by X-ray photoelectron spectroscopy. Our results expand and complete the row of lead chalcogenide-based 2D NPLs, opening up new ways for further pushing the optical properties of 2D NPLs into the infrared and toward technologically relevant wavelengths.

INTRODUCTION

In pursuit of small band gap colloidal semiconductors to cover absorption and emission beyond visible wavelengths, two-dimensional (2D) lead chalcogenide nanosheets (NSs) and nanoplatelets (NPLs) have attracted considerable attention as candidates for solution processable optoelectronics.[1–5] The materials exhibit tunable optical properties, determined primarily by their quantum-confined thickness, which span near-infrared wavelengths (NIR, 750 – 1400 nm (1.65 – 0.89 eV)) and reach up to the short-wave-infrared (SWIR, 1400 – 3000 nm (0.89 – 0.41 eV)).[6–10] Access to these spectral windows is desirable for fiber optic applications, as glass fibers exhibit negligible optical attenuation in this range.[11] The strong vertical confinement in the 2D geometry of NSs and NPLs gives rise to interesting photophysics, for example, efficient charge carrier multiplication (up to 90 % for 4 nm thick PbS NSs, which means that almost all excess photon energy is converted into additional electron-hole pairs),[12] thickness dependent Rashba-type spin-orbit coupling,[13] and highly mobile charge carriers (550 – 1000 cm$^2$V$^{-1}$s$^{-1}$ for PbS NSs with a thickness of 4 – 16 nm).[14]

Consequently, colloidal 2D lead chalcogenides are intriguing as active materials in solar cells,[3] photodetectors,[2] and field effect transistors[15] and have received increasing attention in the realm of single photon emission, which is highly desirable for fiber-based photonic quantum technologies operating at telecommunication wavelengths.[16,17] Noteworthy, PbTe in particular has additionally been thoroughly investigated as an intermediate-temperature thermoelectric material (bulk figure of merit $ZT$ of 0.4 - 0.45 at 300 K) and PbTe quantum wells grown *via* molecular beam epitaxy have shown $Z_{2D}T > 1.2$ due to the high density of states in the 2D geometry.[18,19] However, the colloidal synthesis of 2D lead chalcogenide NSs and NPLs is challenging since anisotropic crystal growth has to be realized for materials with an isotropic crystal structure (rock salt for PbX (X = S, Se, Te)).[1,20–22] While there are several ways to synthesized PbS NPLs[10,23] and NSs[1,4,20,24], the selenium, and especially tellurium analogs, are far less accessible compared to PbS. Syntheses of 2D PbS and PbSe NPLs are commonly conducted in amine solvents (e.g. *n*-octylamine) at comparatively low reaction temperatures (0 – 40 °C). With these methods, the organometallic precursors can nucleate within soft lamellar amine bilayer templates supporting the 2D growth toward NPLs.[20] Following this approach, the limited synthetic access to PbSe and PbTe NPLs is presumably partially caused by a limited variety of precursors which are sufficiently reactive at low temperatures, combined with a typically low stability of the final NPL product under ambient conditions. In addition to autoclave syntheses of PbTe NSs (with a thickness between 20 – 80 nm) by Zhu *et al.*[25] and PbSe/PbTe micropeonies composed of NPLs by Jin *et al.*[26], Chatterjee *et al.*[27] have synthesized $Pb_mBi_{2n}Te_{3n+m}$ NSs with a thickness of a few nanometers *via* a solution-based method.

However, to the best of our knowledge, up to now, no direct colloidal synthesis of ultrathin and optically active 2D PbTe NPLs has been reported.

Concerning the precursor chemistry, thiourea and selenourea as well as their substituted derivatives have been thoroughly studied in the context of nanocrystal (NC) syntheses and have been successfully used, amongst other chalcogen sources, for the direct low-temperature synthesis of PbS and PbSe NPLs resp.[6,20,28,29] Morrison *et al.* obtained 1 nm thick PbS quantum platelets using thiourea as sulfur source at 40 °C and established the lamellar mesophase template mechanism for the formation of the platelets.[20] Similarly, our group has recently applied selenourea to directly synthesize PbSe NPLs at 0 °C with efficient NIR emission (PL quantum yields (PLQY) up to 61 % for PL at 989 nm (1.25 eV) and 27 % for PL at 1265 nm (0.98 eV)).[6,9] However, reports on tellurourea and its derivates are scarce since it is prone to de-telluration, is light and moisture sensitive and hardly soluble in low polarity organic solvents due to elaborate aromatic carbene ligands being required to stabilize the C = Te bond.[30] Consequently, a different precursor chemistry is needed for the synthesis of PbTe NPLs. A common telluride source is trioctylphosphine telluride (TOP-Te), which has been used by Izquierdo *et al.* to synthesize NIR emitting HgTe NPLs (PLQY of 10 % at 880 nm (1.41 eV) with 40 nm (57 meV) FWHM *via* cation exchange from CdTe NPLs).[31] However, TOP-Te lacks reactivity at the low temperatures required to synthesize 2D PbTe NPLs *via* a soft template mechanism.[32–34] Sun *et al.* have used tris(dimethylamino)phosphine telluride as a more reactive alternative to well established TOP-Te sources for the synthesis of CdTe NPLs, (CdTe)$_{13}$ nanoclusters and CdTe nanowires and concluded that transaminated derivates are the actual reactive tellurium source in primary amine solvents.[32]

Here, we adapt aminophosphine precursor chemistry to present a straightforward synthesis for colloidal 2D PbTe NPLs with tunable near infrared PL (910 – 1460 nm (1.36 – 0.85 eV), PLQY 1 - 14 %) by using lead oleate and transaminated tris(dimethylamino)phosphine telluride at low reaction temperatures of 0 °C. The colloidal stability of the as-synthesized PbTe NPLs is increased by quenching the reaction and passivating the NPLs surface with lead iodide in one step.

EXPERIMENTAL SECTION

**Chemicals.** Acetonitrile (≥ 99.5 %), chloroform-d (CDCl$_3$, 99.8 atom % D), isopropanol (≥ 99.5 %), lead(II) oxide (≥ 99.99 %), methanol (≥ 99.8 %), *n*-octylamine (99 %), tellurium powder (30 mesh, 99.99 %), tetrachloroethylene (TCE, ≥ 99 %), triethylamine (≥ 99 %), trifluoroacetic acid (99 %), trifluoracetic anhydride (≥ 99 %), and tris(dimethylamino)phosphine (P(N(CH$_3$)$_2$)$_3$, 97 %) were purchased from Sigma-Aldrich/Merck. *n*-Hexane (97 %) was purchased from Acros Organics. Lead(II) iodide (99.99 %) was purchased from Alfa Aesar. Oleic acid (90 %) was purchased from ABCR. *n*-Octylamine and oleic acid were degassed by the freeze-pump-thaw method three times prior to being stored and used inside a nitrogen-filled glovebox. All other reagents were directly used as received from the listed suppliers.

**Preparation of the Lead Oleate Precursor.** Lead oleate was synthesized following an established procedure by Hendricks *et al.*[28] Accordingly, lead(II) oxide (10 g, 44.8 mmol) was dispersed in acetonitrile (20 ml) at 0 °C; trifluoroacetic anhydride (6.2 ml, 44.8 mmol) and trifluoroacetic acid (0.7 ml, 9.1 mmol) were added, and after 15 min at 0 °C, the clear solution was allowed to heat to room temperature. In a separate flask, oleic acid (90 %, 25.4 g, 81 mmol), triethylamine (14.1 ml, 101 mmol), and isopropanol (180 ml) were mixed. Upon combining the two solutions, lead oleate was immediately obtained as a white precipitate. After recrystallization from isopropanol and generous washing with methanol, the lead oleate was dried under reduced pressure and stored under inert gas conditions ($N_2$) at -25 °C. A stock solution (0.5 M) of lead oleate (2.31 g, 3.00 mmol) in *n*-octylamine (6.0 ml) was prepared and stored inside a nitrogen-filled glovebox and used for multiple PbTe NPL syntheses (within up to one month).

**Preparation of the Aminophosphine Telluride Precursor.** The aminophosphine telluride precursor preparation was adapted from Sun *et al.*[32] with the major difference that an *ex situ* transamination of tris(dimethylamino)phosphine with *n*-octylamine was performed prior to the PbTe NPL synthesis. A precursor stock solution (0.5 M) was prepared by heating a stirred mixture of tellurium powder (255 mg, 2.00 mmol), tris(dimethylamino)phosphine (2 ml, 11.0 mmol), and *n*-octylamine (2 ml, 12.1 mmol) to 100 °C for 3h under inert gas conditions. Subsequently, the ocher colored solution was filtered using a polytetrafluoroethylene syringe filter (pore size 0.2 μm), allowed to cool to room temperature, and used for multiple PbTe NPL syntheses (within up to two weeks).

**Preparation of the Lead Iodide Surface Passivation.** A lead iodide stock solution (0.1 M) for quenching and passivating crude PbTe NPLs was prepared by dissolving lead iodide (461 mg, 1.00 mmol) in *n*-octylamine (7.87 ml, 47.5 mmol) and oleic acid (2.13 ml, 6.75 mmol) at 35 °C for 1 h under stirring inside a nitrogen-filled glovebox.

**PbTe NPL Synthesis.** In a 5 ml round-bottom flask, 0.4 ml of the lead oleate precursor solution described above was diluted in hexane (3.6 ml) and cooled down to 0 °C under stirring. Subsequently, 0.2 ml of the aminophosphine telluride precursor solution were rapidly injected, causing the colorless solution to turn light brown. After a reaction time of 30 min, the increasingly dark mixture was quenched by injecting 2 ml of the lead iodide solution and diluted with hexane (6 ml). The passivated PbTe NPLs were stored at -25 °C inside a nitrogen-filled glovebox.

**NIR PL and Ultraviolet (UV)-Visible (Vis)-NIR Absorbance Spectroscopy.** All samples for optical spectroscopy were prepared by diluting the colloidal PbTe NPLs in TCE (2.5 mL) in a quartz cuvette (quartz glass high performance QS 200 – 2500 nm (6.20 – 0.5 eV) by Hellma) with a path length of 1 cm (optical density below 0.2 at 500 nm (2.48 eV)). NIR PL spectra were acquired using an Edinburgh FLS 1000 UV-Vis-NIR PL spectrometer and a PTI QuantaMaster QM4 spectrofluorometer. The Edinburgh FLS 1000 UV-Vis-NIR spectrometer is equipped with a 450 W ozone free xenon arc lamp for excitation. PL was monitored using a liquid nitrogen cooled InGaAs NIR photomultiplier tube 1650 detector from Edinburgh. The PTI QuantaMaster QM4 spectrofluorometer is equipped with a 75 W steady-state xenon short arc lamp for excitation. PL was monitored using a liquid nitrogen cooled PTI P1.7R detector module (Hamamatsu PMT R5509-72). To avoid higher order excitation light, a RG780 long pass filter glass plate (thickness 3 mm) was used in the emission path. Spectral selection was achieved by single grating monochromators (excitation: 1200 grooves/mm, 300 nm blaze; NIR emission: 600 grooves/mm, 1200 nm blaze). PL spectra were collected by exciting PbTe NPLs at 500 nm (2.48 eV). Absolute PLQYs were determined with the FLS 1000 spectrometer using an integrating sphere. For this, scattering at 500 nm (2.48 eV) and the PL in the NIR region of both, the pure solvent and the NPLs, were separately measured, considering the sensitivity difference of both detectors with a correction factor. Multichannel scaling (MCS) PL lifetime analysis was performed with the FLS 1000 spectrometer equipped with a picosecond pulsed diode laser (pulse width of 110 ps at 445.1 nm (2.79 eV)) from Edinburgh Instruments. Vis-NIR absorbance spectra were collected using a Cary 5000 spectrophotometer from Agilent Technologies.

**High Resolution Transmission Electron Microscopy (HRTEM).** HRTEM images were obtained using a JEOL ARM 200F equipped with a CETCOR aberration corrector operating at 80 kV. Samples for TEM analysis were prepared by drop casting the colloidal PbTe NPL solution onto graphene-coated Quantifoil grids acquired from Graphenea. The dry samples were washed with hexane.

**Powder X-ray Diffraction (PXRD).** PXRD patterns were measured in Bragg-Bretano geometry with a Bruker D8 Advance equipped with a Cu $K_{\alpha_1}$ source operating at 40 kV and 30 mA. For measuring diffractograms, a colloidal PbTe NPL solutions were prepared by drop casting onto single crystal silicon sample holders.

**Grazing-Incidence Wide Angle X-ray Scattering (GIWAXS).** GIWAXS data were obtained at the P03 beamline of the PETRA III synchrotron facility at an incidence angle of 0.4° with an X-ray photon energy of 11.875 keV. The diffraction data were recorded with a Lambda 9M detector placed 205 mm behind the sample. The samples were prepared by drop casting the colloidal PbTe NPL solution onto silicon wafers (5 mm x 5 mm, p-type doped with boron, <100> surface, purchased from Plano).

**Nuclear Magnetic Resonance (NMR) Spectroscopy.** NMR measurements were conducted with a Bruker Avance III HDX 400 with a frequency of 400 MHz. Samples were prepared by filtering each specimen through a polytetrafluoroethylene syringe filter (pore size 0.2 μm) and diluting in dry chloroform-d in an NMR tube inside a nitrogen-filled glovebox. Spectra were analyzed using the Bruker TopSpin 4.2.0 software. The residual solvent peak of chloroform-d (7.26 ppm) was used as an internal reference for the $^1$H spectra. The $^{31}$P spectra were measured with proton decoupling and indirectly referenced to the $^1$H NMR frequency of the same sample using the "xiref"-function of Bruker TopSpin 4.2.0.

**X-ray Photoelectron Spectroscopy (XPS).** XPS data were collected using a PHI 5000 VersaProbe III from ULVAC-PHI. The spectra were obtained with an aluminum X-ray source (Al $K_\alpha$ = 1486.6 eV) operating at 24.4 W with a beam diameter of 100 μm. Survey spectra were measured with a pass energy of 224 eV; high-resolution spectra were acquired with a pass energy of 27 eV. Charging effects were accounted for by setting the C 1s peak of sp$^3$ carbon to 284.4 eV. The samples for XPS analysis were prepared by drop casting the colloidal PbTe NPL solution onto a silicon wafer from Plano and drying under vacuum overnight.

RESULTS AND DISCUSSION

**Tunable optical properties of PbTe NPLs.** Figure 1a-c includes HRTEM images of PbTe NPLs with three different lateral size distributions (corresponding size histograms are depicted in Figure S1, typical TEM images used for determining the size are shown in Figure S2). The majority of NPLs resemble a rectangular shape with rounded edges with occasionally more irregular shapes as well. Their crystal phase is determined by selected area electron diffraction (SAED, see Figure 1d) and PXRD (see Figure 1e). The SAED pattern is composed of the characteristic diffraction peaks of the cubic PbTe phase (space group $Fm\bar{3}m$) with the corresponding lattice spacings of 3.23 Å (200), 2.30 Å (220), and 1.50 Å (331) (PDF card 01-072-6645). Complimentary to SAED, PXRD underpins the cubic rock salt structure of PbTe for the whole NPL ensemble. Additional information on the size and 2D shape of PbTe NPLs is obtained by GIWAXS measurements of drop casted solutions on silicon substrates (see Figure 1f). Clearly visible diffraction peaks suggest that the NPLs form a superlattice. The significant difference in the spacing between the diffraction peaks in vertical and horizontal directions indicates that the superlattice has substantially different lattice parameter values in these two directions. Such a difference can be explained by the highly anisotropic shape of the NPLs. A possible superlattice structure is shown in of Figure 1g (see Figure S3 for complementary X-ray reflectivity data). The small interparticle distance of 0.5 nm suggests that the ordered NPLs (in this measurement) consist of a single (001) atomic layer of PbTe separated by iodide ligands (Pb-I bond length of ~0.32 nm).[35]

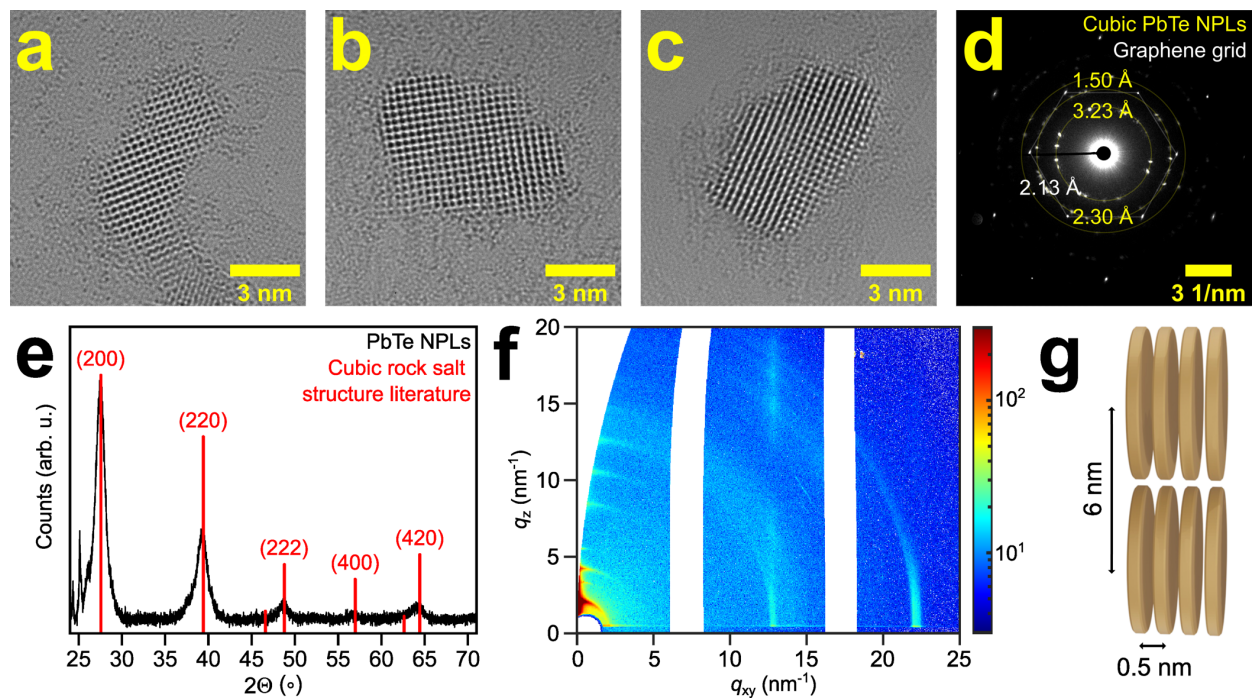

**Figure 1.** (a-c) HR-TEM images of PbTe NPLs exhibiting a slightly rectangular shape. (d) SAED of the PbTe NPL sample shown in (a) reveals their cubic rock salt crystal structure. (e) PXRD of a macroscopic PbTe NPL ensemble exhibits the cubic rock salt structure. (f) GIWAXS diffraction pattern of drop casted PbTe NPLs on a silicon substrate (white areas correspond to detector gaps). The diffraction peaks in vertical direction correspond to stacking in real space with a period of ~6 nm. The diffraction peaks in horizontal direction indicate a periodicity of ~0.5 nm. (g) Possible scheme for a superlattice formed by PbTe NPLs with corresponding length scales.

Figure 2a shows the absorbance and PL spectrum of PbTe NPLs exhibiting a weakly pronounced feature at 817 nm (1.52 eV) and associated PL at 1075 nm (1.15 eV) with a FWHM of 160 meV (150 nm) and a PLQY of 11 %. Besides the highly confined thickness of the NPLs (see Figure S4), the large exciton Bohr radius of PbTe ($a_{B,transversal}$ = 152 nm, $a_{B,longitudinal}$ = 12.9 nm)[36] results in an additional strong confinement in their lateral dimension (i.e. strong confinement in all three dimensions, where $r/a_B \ll 1$).[37] Changes in the thickness and lateral size of the PbTe NPLs consequently lead to a continuously shift of the PL position and are synthetically achieved by altering the ratio of Pb:Te used for synthesis. With a higher lead excess during synthesis, larger PbTe NPLs with PL shifted further into the NIR are obtained (see Figure S5). Figure 2b illustrates the tunability of the PL maximum of PbTe NPLs in the range of 910 - 1460 nm (1.36 – 0.85 eV). The wide range over which the PbTe NPLs PL can be tuned, combined with the absence of discrete steps (as observed for e.g. CdSe NPLs),[38,39] point toward a combined role of thickness and lateral size of the NPLs in shifting the PL maxima. PbTe NPLs shown in Figure 1a-c with average lengths of 7.1 ± 0.9 nm, 7.7 ± 1.0 nm, and 8.7 ± 2.2 nm resp., exhibit corresponding PL at 1217 nm (1.02 eV), 1293 nm (0.96 eV), and 1459 nm (0.85 eV) (see Figure S1 and S5). NPL widths do not strictly follow this trend (see Figure S1), underpinning that the most confined thickness dimension predominately determines the PL position, while the lateral dimensions play minor role (it has to be noted that the determined NPL widths exhibit higher relative error margins due to the varying NPL shape). The FWHM of the PL signals shown in Figure 2b varies between 282 meV (190 nm) for PL at 910 nm (1.36 eV) and 100 meV (173 nm) for PL further in the NIR at 1460 nm (0.85 eV). Not all PL signals strictly follow this trend, which we attribute to an inhomogeneous broadening of the PL signals, caused by the size distribution of the NPLs. Compared to FWHM values of directly synthesized colloidal 2D PbSe NPLs (214 meV (164 nm) for PL at 976 nm (1.27 eV) and 184 meV (269 nm) for PL at 1362 nm (0.91 eV))[6], the PbTe NPLs shown here exhibit broader PL at shorter wavelengths, but significantly narrower PL at longer wavelengths approaching the low-loss third telecommunication window around 1550 nm (0.80 eV).[11] Within the PL tunability range, the PLQY of PbTe NPLs gradually rises from 1 % near 900 nm (1.38 eV) up to 15 % for PL maxima above 1265 nm (0.98 eV) (Figure S6). This trend of a generally increasing PLQY towards longer NIR wavelengths is different from previous descriptions for PbS and PbSe NCs and NPLs, where larger radii or lateral sizes are associated with a reduced PLQY.[6,9,40] Here, a decreasing PLQY in smaller band gap NCs and NPLs is explained by the *energy gap law*, which expresses the exponential relationship of transition rate and energy difference between two states: Non-radiative recombination rates are increased in larger NCs and NPLs, where dark trap states are located closer to the band edge.[40] While trap states in PbX NPLs are expected to be caused mainly by undercoordinated edge or ledge atoms at the surface (of which larger NPLs contain a lower fraction compared to smaller NPLs), the *energy gap law* counteracts this positive influence on the PLQY for PbS and PbSe NPLs. In contrast, for PbTe, the lower fraction of edge or ledge atoms in larger NPLs seems to outweigh the negative effect of the *energy gap law* on the PLQY. This reversed influence in PbTe NPLs could be e.g. caused by dark trap states located further away from the conduction band edge energetically compared to PbS and PbSe NPLs and reducing the effect of accelerated recombination to dark states in larger NPLs.

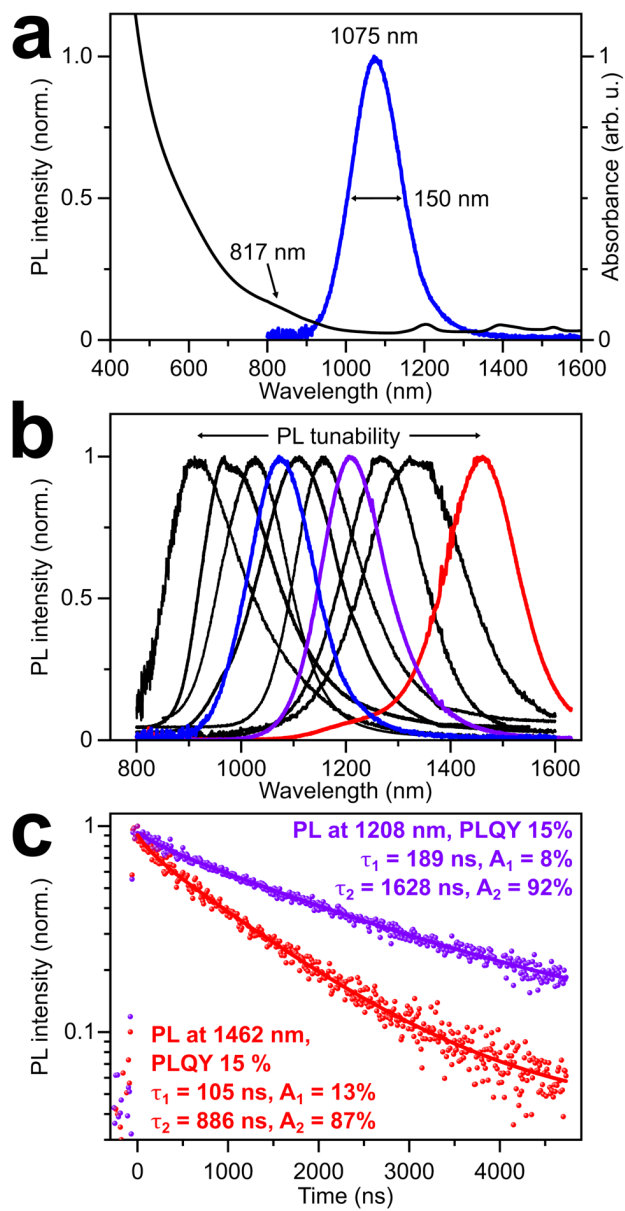

**Figure 2.** (a) Vis-NIR absorbance, (a, b) NIR PL spectra and (c) MCS lifetimes of colloidal 2D PbTe NPLs. (a) Exemplary PbTe NPLs exhibit a length and width of (6.9 ± 0.75) x (4.9 ± 0.9) nm², excitonic absorbance at 817 nm (1.52 eV), PL at 1075 nm (1.15 eV) with a FWHM of 160 meV (150 nm) and a PLQY of 11 %. (b) The PL of PbTe NPLs is tunable between 910 - 1460 nm (1.36 – 0.85 eV) by adjusting the NPLs size *via* the reaction

conditions. (c) MCS lifetimes of PbTe NPLs with PL at 1208 nm (1.03 eV) and 1462 nm (0.85 eV), resp. The biexponential decay is dominated by a long lifetime contribution ($\tau_{2,1208\,nm}$ = 1628 ns and $\tau_{2,1462\,nm}$ = 886 ns) for both PbSe NPL samples and attributed to band edge PL.

MCS PL lifetimes of PbTe NPLs (see Figure 2) are fitted biexponentially with a minor short life time contribution and a dominating long lifetime contribution. For PbTe NPLs emitting at 1208 nm (1.03 eV) and 1462 nm (0.85 eV), resp., time constants of $\tau_{1,1208\,nm}$ = 189 ns (8 %) and $\tau_{2,1208\,nm}$ = 1628 ns (92 %) as well as $\tau_{1,1462\,nm}$ = 105 ns (13 %) and $\tau_{2,1462\,nm}$ = 886 ns (87 %) are determined ($R^2$ = 0.99 and $R^2$ = 0.99). In accordance with previous work on PbX (X = S, Se) NCs and NPLs, we attribute these two lifetime contributions to two different radiative recombination pathways. We assign the shorter lifetime contribution to the *energy gap law* and associated defect emission, while the dominant longer lifetimes are ascribed to band edge emission.[9,41,42] The biexponential nature and the depicted lifetimes are in good agreement with values reported by Lin *et al.* for spherical PbTe QDs capped with oleylamine in TCE ($\tau_1$ = 96 ns (34 %) and $\tau_2$ = 1490 ns (66 %)),[43] as well as our previous work on colloidal PbSe NPLs passivated with CdCl$_2$ ($\tau_{1PL}$ = 168 ns (15 %) and $\tau_{2PL}$ = 1320 ns (85 %) for PL at 980 nm(1.27 eV)).[6] The smaller contribution of the shorter lifetime in PbI$_2$ passivated PbTe NPLs (8 % and 13 % resp.) (and the CdCl2 passivated PbSe NPLs (15 %)) compared to the 34 : 66 ratio for oleylamine capped PbTe QDs, indicate that the fast recombination can be suppressed by a more complete surface passivation, i.e. it is related to defect emission. Furthermore, Galle *et al.* determined slightly longer PL lifetimes of 0.7 – 2.7 µs for PbSe NPLs synthesized *via* cation exchange from CdSe (depending on the thickness of the initial CdSe NPLs)[6,8] and Babaev *et al.* reported PL lifetimes around 2 µs for PbSe/PbS core/shell NPLs obtained *via* cation exchange from CdSe/CdS core/shell NPLs.[44] It should be noted that Khan *et al.* found significantly shorter PL lifetimes in the range of 8.4 - 59 ns for colloidal 2D PbS NPLs with PL ranging from 735 - 748 nm, although such short lifetimes have not been reported for selenium (and tellurium) analogs.[23] Generally, room temperature PL lifetimes of lead chalcogenide NPLs are significantly longer in comparison to CdSe NPLs (faster than 10 ns).[38,39] Theoretical calculations for spherical PbSe and CdSe NCs have ascribed long PL lifetimes in PbSe NCs to a reduced oscillator strength of the lowest-energy bright exciton states, caused by intervalley coupling among the four equivalent L points in the Brillouin zone.[45] Given that PbTe exhibits the same crystal structure (cubic rock salt) as well as a similar band gap and structure as PbSe[46], we assume the same holds for the comparatively long PL lifetimes in 2D PbTe NPLs.

**Synthesis of PbTe NPLs.** Lead oleate and lead acetate (which is converted to lead oleate *in-situ*) are commonly used as lead source for the synthesis of 2D PbS and PbSe NPLs.[6,20] Lead oleate was also chosen as Pb precursor in this work, for PbTe NPLs, as it is readily soluble in amines at low temperatures. Our initial efforts for finding a suitable tellurium precursor are based on work by Sun *et al.*, who used tris(dimethylamino)phosphine telluride for the synthesis of CdTe NPLs, (CdTe)$_{13}$ nanoclusters and CdTe nanowires at temperatures as low as 70 °C in case of the nanoclusters.[32] For this, tellurium granules were dissolved in tris(dimethylamino)phosphine at a molar ratio of 1:5.5 at 100 °C for 3 h, and yielded a yellow solution. However, when cooling down to room temperature, white crystals of TeP(N(CH$_3$)$_2$)$_3$ precipitated from solution, hindering the use in a low temperature synthetic approach. When circumventing the precipitation by injecting the yellow TeP(N(CH$_3$)$_2$)$_3$ solution at elevated temperatures prior to precipitation, no reaction with Pb(oleate)$_2$ at 0 °C occurred. This observation can be rationalized by findings of Sun *et al.* who inferred that transaminated TeP(N(CH$_3$)$_2$)$_3$ derivates are the active tellurium precursor in their syntheses. The transamination of aminophosphines with amines, usually oleyl- or octylamine, is well known.[47,48] For instance, Tessier *et al.* reported on the role of tris(oleylamino)phosphine during the synthesis of InP quantum dots in oleylamine as an explanation for similar results regardless of the initial use of tris(dimethylamino)phosphine or tris(diethylamino)phosphine.[49]

For the synthesis of PbTe, we apply *in-situ* transamination by dissolving tellurium in a mixture of tris(dimethylamino)phosphine and *n*-octylamine at 100 °C for 3 h prior to the PbTe NPL synthesis (see Figure 3a). At first, we obtained the same yellow solution as in the absence of the octylamine, but over time the color of the mixture changed from yellow to ocher and no precipitation occurred when cooling down to room temperature. During the reaction time, intense gas formation was observed, similarly to what has been described by Tessier *et al.* (via a CuSO$_4$ gas trap)[49] and Sun *et al.* (*via* $^{13}$C{$^1$H} and $^1$H NMR)[32] as dimethylamine and indicating a successful transamination reaction. Figure 3b and c show $^1$H- and $^{31}$P{$^1$H}-NMR spectra of pure tris(dimethylamino)phosphine (P(N(CH$_3$)$_2$)$_3$ (1)), tris(dimethylamino)phosphine telluride (TeP(N(CH$_3$)$_2$)$_3$ / P(N(CH$_3$)$_2$)$_3$ (2)), and transaminated tris(dimethylamino)phosphine telluride (TeP(N(CH$_3$)$_2$)$_{3-x}$(NHR)$_x$ / P(N(CH$_3$)$_2$)$_{3-x}$(NHR)$_x$, R = C$_8$H$_{17}$ (3 - 5)) in CDCl$_3$ in the range of 0.7 - 3.05 ppm and 95 - 135 ppm (0 – 95 ppm available in the SI (Figure S7)), resp. The $^{31}$P{$^1$H}-NMR spectra were evaluated by comparison to literature values and with the help of heteronuclear multiple bond correlation (HMBC) NMR spectra shown in the SI (Figure S8-10).[32,49,50] The $^1$H-NMR spectrum of pure tris(dimethylamino)phosphine (1) features a single doublet resonance at 2.48 ppm with a coupling constant of 9.15 Hz corresponding to the PN*CH$_3$* protons coupling to the phosphorus atom. The associated $^{31}$P{$^1$H}-NMR spectrum shows a single sharp resonance of *P*(N(CH$_3$)$_2$)$_3$ at 122 ppm. Upon telluration (2) strong broadening of the *P*(N(CH$_3$)$_2$)$_3$ resonance around 122 ppm occurs, which might be attributed to the enduring reaction dynamic. The PN*CH$_3$* resonance in the $^1$H-NMR is barely shifted to 2.47 ppm with an increased coupling constant of 9.69 Hz. The $^1$H-NMR spectra of transaminated tris(dimethylamino)phosphine telluride (TeP(N(CH$_3$)$_2$)$_{3-x}$(NHR)$_x$ / P(N(CH$_3$)$_2$)$_{3-x}$(NHR)$_x$, R = C$_8$H$_{17}$ (3 - 5)) exhibits the C*H$_3$* resonance of P(N(CH$_3$)$_2$)$_3$ (at 2.48 ppm, *J* = 9.55 Hz) as well as several multiplets or broad signals attributed to the protons of the aliphatic carbon chain of *n*-octylamine or *n*-octylamino groups in different chemical environments. The corresponding $^{31}$P{$^1$H}-NMR contains three new resonances upfield shifted with respect to the tris(dimethylamino)phosphine telluride resonance at 122 ppm. The three resonances at 110 ppm, 112 ppm and 114 ppm are assigned to the triple (5), double (4) and single (3) transaminated aminophosphine thus confirming the successful transamination of tris(dimethylamino)phosphine telluride.[32,49] This allocation is substantiated by $^{31}$P{$^1$H}-NMR spectra of aliquots from the transamination reaction mixture at different temperatures (Figure S11a). The three $^{31}$P resonances arise in order from 3 to 5 when heating up from 40 °C to the reaction temperature of 100 °C, indicating a gradual substitution of the initial dimethylamino group (-NMe$_2$). Combined with the images of the collected aliquots (Figure S11b), showing a distinct color change upon reaching a 100 °C, these spectra underpin the prerequisite of the *ex situ* transamination, as substitution would not occur at the PbTe NPL synthesis temperature of 0 °C.

By utilizing the *ex-situ* transaminated aminophosphine telluride for our synthesis, PbTe NPLs were obtained within a reaction time of 30 min at 0 °C. To conclude a specific reaction pathway and a definitive reason for the necessity of the transamination, a comprehensive mechanistic study would be required. In analogy to Sun *et al.*, we expect a nucleophilic substitution of one oleate (in lead oleate) by the transaminated aminophosphine telluride *via* a nucleophilic attack from the telluride on the lead, followed by a nucleophilic attack of the eliminated oleate on the phosphorous, which results in the formation of PbTe and the second oleate (see Figure S12).[32] Coherently, we assume the enhanced reactivity of the transaminated aminophosphine telluride may be caused by the increased nucleophilicity of the phosphorous atom. The electron-donating character of the long alkyl chains combined with the electron-donating character of the amino nitrogen lone pairs contributes to the stabilization of a positive charge on the phosphorous atom and facilitates the release of Te$^{2-}$ for the formation of PbTe.[32,49] In contrast to high-temperature syntheses with aminophosphine precursors in which the high-boiling amine solvents primarily determine the reactivity, the *ex-situ* transamination allows to further tune the precursor reactivity *via* the organic tail group of the amine (see Figure S13). The use of aminophosphine telluride precursors transaminated with linear aliphatic primary amines of increasing chain length for PbTe NPL synthesis results in a bathochromic shift of the PL maximum. We propose that longer alkyl chains reduce the reactivity of the tellurium precursor due to steric hinderance and consequently result in the formation of larger PbTe NPLs with PL further in the NIR.

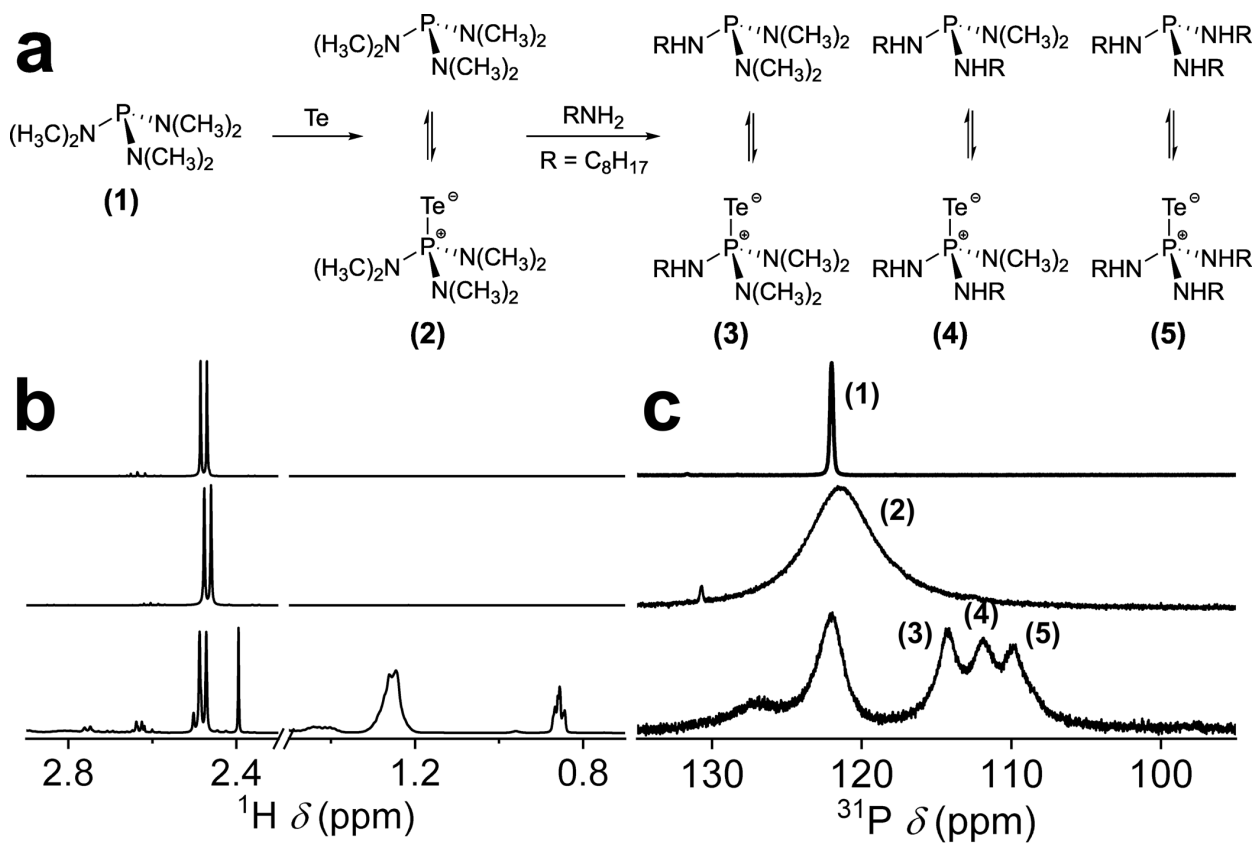

**Figure 3.** (a) Scheme of the combined telluration and transamination reaction of tris(dimethylamino)phosphine (P(N(CH3)2)3) with tellurium and *n*-octylamine. (b) $^1$H- and (c) $^{31}$P{$^1$H}- NMR spectra of pure tris(dimethylamino)phosphine (P(N(CH$_3$)$_2$)$_3$ (1)), tris(dimethylamino)phosphine telluride (TeP(N(CH$_3$)$_2$)$_3$ / P(N(CH$_3$)$_2$)$_3$ (2)), and transaminated tris(dimethylamino)phosphine telluride (TeP(N(CH$_3$)$_2$)$_{3-x}$(NHR)$_x$ / P(N(CH$_3$)$_2$)$_{3-x}$(NHR)$_x$, $R = C_8H_{17}$ (3 - 5)) in CDCl$_3$. The presence of three phosphorous resonances corresponding to the triple (5), double (4), and single (3) transaminated aminophosphine underpins the successful transamination reaction.

**Passivation of PbTe NPLs.** To increase the colloidal stability of the as-synthesized PbTe NPLs, post-synthetic surface passivation with lead iodide was performed. Lead iodide can act as both, X- and Z-type ligand and passivate dangling bonds of unsaturated lead or chalcogen surface sites.[9] Figure 4 shows XPS analysis of the Pb-4f and Te-3d core level regions of pristine and PbI$_2$ passivated PbTe NPLs. All components are fitted by symmetric Voigt functions (convolution of a Lorentzian- and Gaussian distribution) with maxima corresponding to the binding energies of the different atomic species. The spin-orbit doublet in the range of 135 – 145 eV is composed of the 4f$_{7/2}$ and 4f$_{5/2}$-signals of lead and is fitted by four different components (Figure 4a and b). The lowest energy component at 136.2 eV and 141.1 eV for pristine PbTe NPLs and 136.3 eV and 141.1 eV for PbI$_2$ passivated PbTe NPLs is attributed to elemental lead in the oxidation state zero. The formation of metallic lead *via* photodegradation under XPS conditions has previously been reported for lead halide perovskite nanocrystals capped with oleic acid/oleylamine.[51,52] In particular, the presence of PbI$_2$/I$^-$ is associated with a high degree of photodegradation, and thus explaining the high intensity of the Pb(0) signal in PbI$_2$ passivated NPLs. The second component at 137.3 eV and 142.1 eV (pristine) as well as 137.0 eV and 141.9 eV (passivated) PbTe NPLs resp., is ascribed to lead bound to tellurium.[53,54] The highest energy component in case of the pristine PbTe NPLs occurs at 138.8 eV and 143.6 eV and corresponds to lead bound to carboxylate.[55,56] Upon surface passivation with lead iodide, this signal disappears and a new component, ascribed to lead bound to iodide, emerges at 137.9 eV and 142.8 eV. We conclude a successful passivation of the pristine PbTe NPL surfaces by X-type ligand exchange of oleate with iodide. The tellurium 3d$_{3/2}$ and 3d$_{5/2}$ spin-orbit doublet in the range of 570 - 584 eV in Figure 4c and d is fitted by a single component assigned to tellurium bound to lead.[57] The small shift of 0.3 eV for the signals of the passivated PbTe NPLs (571.8 eV to 572.1 eV and 582.2 eV to 582.5 eV, resp.) could be an indication of the change in the chemical environment upon introduction of the iodide to the NPLs surface. Figure S14 shows a photograph of the pristine PbTe NPLs compared to the passivated NPLs 24 h after the synthesis. Pristine NPLs exhibited a color change from dark brown to black, and a black precipitate formed in solution, while metallic/elemental tellurium formed at the inside of the vial. On the other hand, iodide passivated PbTe NPLs preserve their dark brown color, colloidal stability and optical properties. We assume a double role of the lead iodide in passivating the NPL surface and in forming a stable iodophosphonium salt $[IP(N(CH_3)_2)_{3-x}(NHR)_x]^+I^-$ ($R = C_8H_{17}$) as a by-product in analogy to the proposed pathway by Sun *et al.* for the reaction of transminated tris(dimethylamino)phosphine with $CdCl_2$ toward CdTe nanostructures.[32] By employing $PbI_2$ passivation, we obtain colloidal PbTe NPL solutions that are long-term stable (multiple months) and can also be post-processed for measurements under ambient conditions and be drop casted or spin coated onto different substrates.

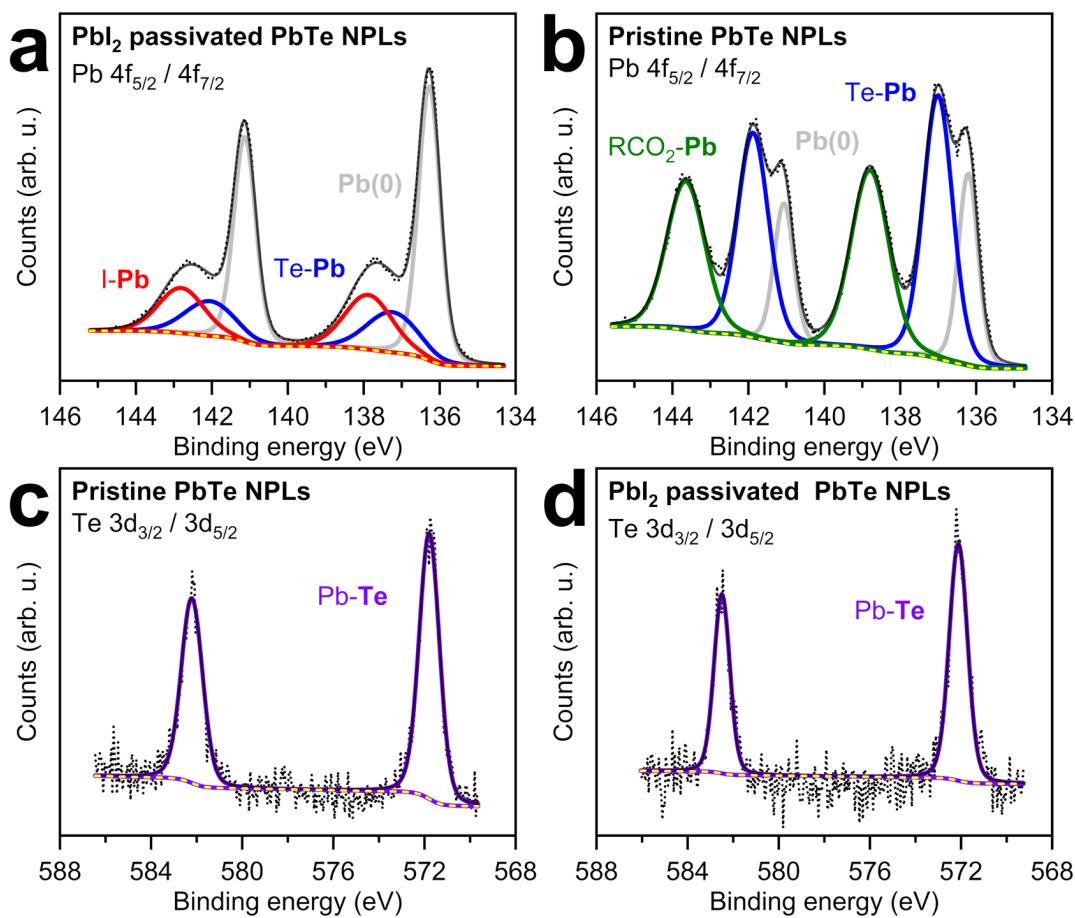

**Figure 4.** XPS analysis of the (a, b) Pb-4f and (c, d) Te-3d core levels of pristine and PbI$_2$ passivated PbTe NPLs. The RCO$_2$-Pb (R = C$_7$H$_{14}$CH=CHC$_8$H$_{17}$) component vanishes upon lead iodide treatment and is replaced by an I-Pb signal, indicating the successful X-type ligand exchange of oleate with iodide. Passivation with lead iodide results in improved colloidal stability compared to pristine PbTe NPLs and leads to preservation of the optical properties during storage of PbTe NPLs (see Figure S14).

CONCLUSION

Up to now, there has been no direct wet-chemical pathway to 2D PbTe NPLs with tunable and efficient NIR photoluminescence. Here, we demonstrate a low temperature colloidal synthesis of 2D PbTe NPLs with cubic rock salt structure by using highly reactive aminophosphine telluride precursor chemistry. Our corresponding comprehensive NMR study shows the synthetic importance of an *ex-situ* transamination reaction of tris(dimethylamino)phosphine telluride for sufficient precursor reactivity at 0 °C to form 2D PbTe NPLs. Associated GIWAXS measurements confirm the 2D geometry of PbTe NPLs and the formation of superlattices from stacked NPLs. Post-synthetically $PbI_2$ passivated PbTe NPLs exhibit tunable NIR PL between 910 – 1460 nm (1.36 – 0.85 eV) (PLQY 1 – 15 %) and narrow FWHM (100 meV (173 nm)), especially at longer wavelengths. By matching the precursor reactivity of aminophosphine telluride with lead oleate, we have now complemented synthetic access to 2D PbS and PbSe NPLs by 2D PbTe, which will help to further tune, explore, and make use of the strong excitonic effects in 2D NPLs in the NIR.

ACKNOWLEDGMENT

We are grateful to Elena Chulanova, Anton Pylypenko, Ingrid Dax, Matthias Schwarzkopf, Dmitry Lapkin and Frank Schreiber for the help with GIWAXS measurements and fruitful discussions. We wish to thank the Deutsches Elektronen-Synchrotron (DESY) in Hamburg for making their accelerator facilities accessible to us by granting the proposal I-20220914 "GIWAXS characterization of colloidal 2D lead chalcogenide nanoplatelets and nanosheets" at beamline PETRA III. Access to the XPS instrument was granted by the Cluster of Excellence PhoenixD under DFG major equipment number 448713396. We are grateful to Armin Feldhoff for providing access to the PXRD facility. L. B., L. F. K. and J. L. gratefully acknowledge funding by the Deutsche Forschungsgemeinschaft (DFG, German Research Foundation) under Germany's Excellence Strategy within the Cluster of Excellence PhoenixD (EXC 2122, Project ID 390833453). J. L. is thankful for additional funding by an Athene Grant of the University of Tübingen (by the Federal Ministry of Education and Research (BMBF) and the Baden-Württemberg Ministry of Science as part of the Excellence Strategy of the German Federal and State Governments).

ASSOCIATED CONTENT

The following files are available free of charge.

Supporting Information: Size histograms of PbTe NPLs with different lateral size distributions; Typical TEM images of PbTe NPLs used for determining lateral NPL dimensions; X-ray reflectivity measurements of PbTe NPLs on a silicon wafer; Plot of the energy of the first excitonic transition *vs.* lateral sizes of 2D PbTe NPLs compared to a sizing curve for spherical PbTe NCs; PL spectra of PbTe NPLs synthesized with three different Pb:Te ratios; Plot of the PLQY vs. the associated PL maxima of PbTe NPLs; $^{31}P\{^{1}H\}$ NMR spectra of pure

tris(dimethylamino)phosphine, tris(dimethylamino)phosphine telluride, and transaminated tris(dimethylamino)phosphine telluride in the range of 0 – 95 ppm; $^{1}$H-$^{31}$P HMBC spectrum of pure tris(dimethylamino)phosphine; $^{1}$H-$^{31}$P HMBC spectrum of tris(dimethylamino)phosphine telluride; $^{1}$H-$^{31}$P HMBC spectrum of transaminated tris(dimethylamino)phosphine; $^{31}$P{$^{1}$H} NMR spectra of aliquots from a tris(dimethylamino)phosphine telluride transamination reaction at different temperatures; Proposed nucleophilic reaction pathway for the formation of PbTe NPLs; PL spectra of PbTe NPLs synthesized using different aminophosphine tellurides; Image of pristine and PbI$_2$ passivated PbTe NPLs 24 h after synthesis.

## AUTHOR INFORMATION

### Corresponding Author

*jannika.lauth@uni-tuebingen.de

### Author Contributions

The manuscript was written through contributions of all authors. All authors have given approval to the final version of the manuscript.

### Notes

The authors declare no competing financial interest.

## ABBREVIATIONS

2D, two-dimensional; FWHM, full width at half maximum; GIWAXS, grazing-incidence wide angle X-ray scattering; HMBC, heteronuclear multiple bond correlation; HRTEM, high resolution transmission electron microscopy; MCS, multichannel scaling; NCs, nanocrystals, NIR, near-infrared; NMR, nuclear magnetic resonance; NPLs, nanoplatelets; NSs, nanosheets; PL, photoluminescence; PLQY, photoluminescence quantum yield; PXRD, powder X-ray diffraction; SAED, selected area electron diffraction; SWIR, short-wave-infrared; TCE, tetrachloroethylene; TOP, trioctylphosphine; UV, ultraviolet; Vis, visible; XPS, X-ray photoelectron spectroscopy.

# Supporting Information

Solving the Synthetic Riddle of Colloidal 2D PbTe Nanoplatelets with Tunable Near-Infrared Emission

*Leon Biesterfeld [a,b,c], Mattis T. Vochezer [a], Marco Kögel [d], Ivan A. Zaluzhnyy [e], Marina Rosebrock [b,c], Lars F. Klepzig [b,c], Wolfgang Leis [f], Michael Seitz [f], Jannik C. Meyer [d,e], Jannika Lauth [*a,b,c]*

a – Institute of Physical and Theoretical Chemistry, Eberhard Karls University of Tübingen, Auf der Morgenstelle 18, D-72076 Tübingen, Germany.

b – Cluster of Excellence PhoenixD (Photonics, Optics, and Engineering – Innovation Across Disciplines), Welfengarten 1A, D-30167 Hannover, Germany.

c – Institute of Physical Chemistry and Electrochemistry, Leibniz University Hannover, Callinstr. 3A, D-30167 Hannover, Germany.

d – Natural and Medical Sciences Institute at the Eberhard Karls University of Tübingen, Markwiesenstr. 55, D-72770 Reutlingen, Germany.

e – Institute of Applied Physics, Eberhard Karls University of Tübingen,

Auf der Morgenstelle 10, D-72076 Tübingen, Germany

f – Institute of Inorganic Chemistry, Eberhard Karls University of Tübingen,

Auf der Morgenstelle 18, D-72076 Tübingen, Germany.

*jannika.lauth@uni-tuebingen.de

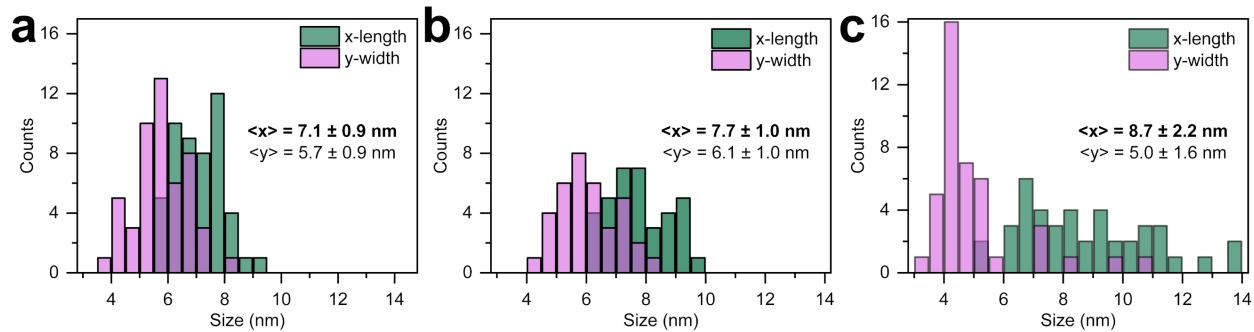

**Figure S1.** Size histograms of the PbTe NPLs shown in the TEM images in Figure 1 of the main manuscript. X-lengths were determined by measuring the longest dimension of the NPLs, y-width is the longest distance orthogonal to the x-length.

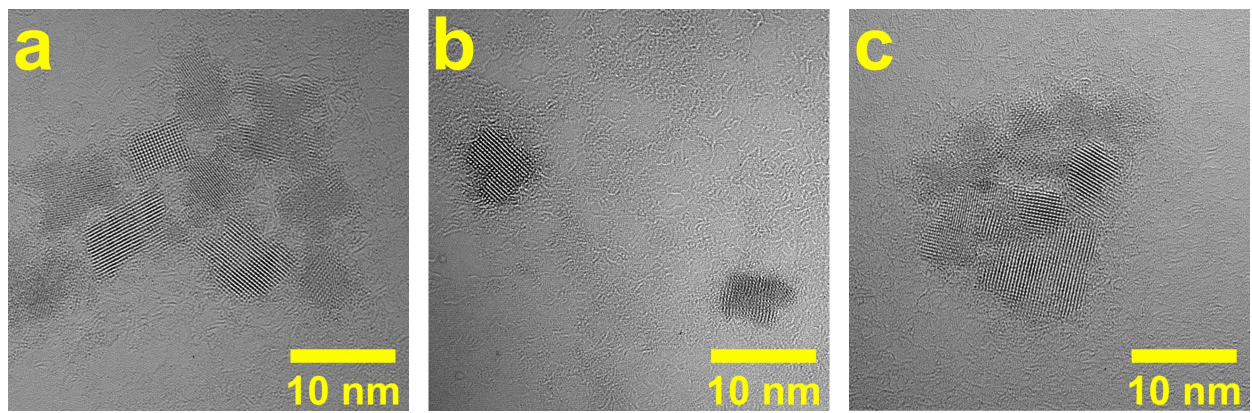

**Figure S2.** Typical TEM images used for determining lengths and widths of PbTe NPLs shown in Figure 1 of the main manuscript.

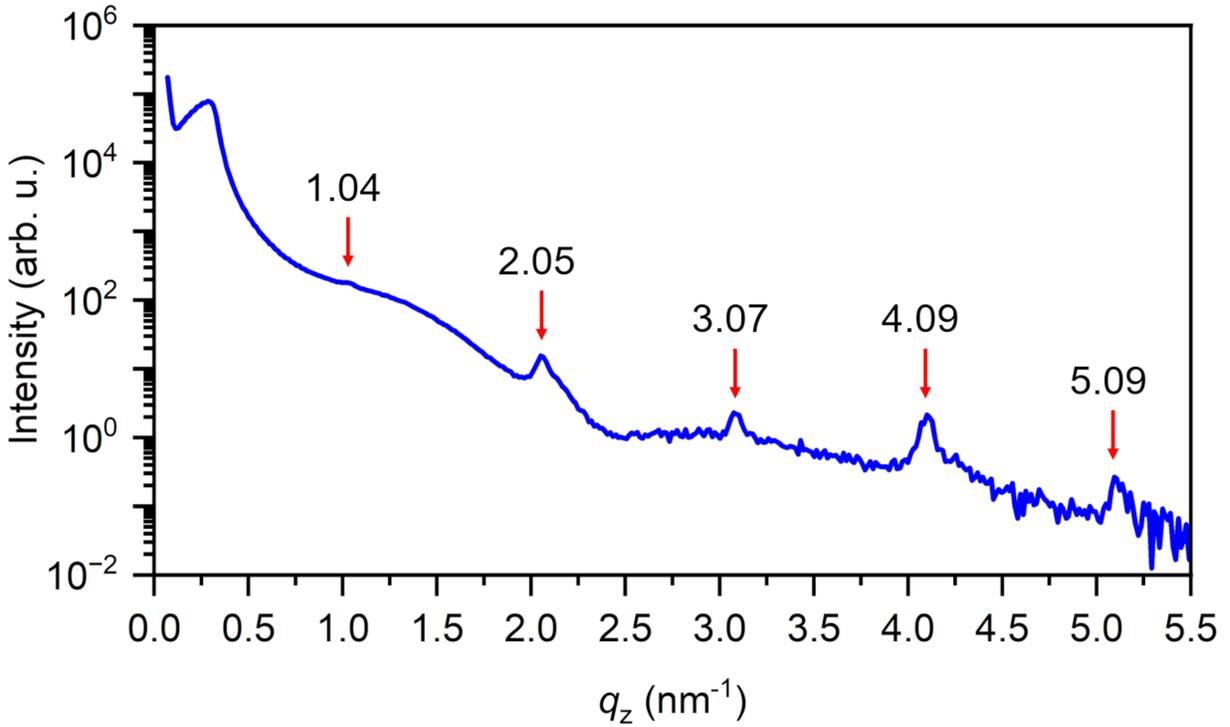

**Figure S3.** Additional information on the vertical stacking of the PbTe NPLs superlattice was obtain by an X-ray reflectivity experiment. The scattering data were recorded using Cu $K_{\alpha_1}$ radiation with an energy of 8.064 keV on a laboratory diffractometer (3303TT, GE). Due to a weak scattering signal, data could only be recorded up to scattering vector $\Delta q_z \approx 6$ nm$^{-1}$. Superlattice peaks with a spacing $\Delta q_z \approx 1$ nm$^{-1}$ are highlighted with red arrows.

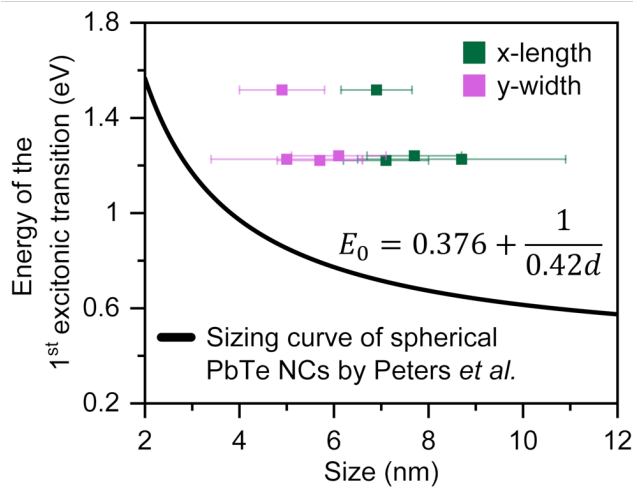

**Figure S4.** Energy of the first excitonic transition *vs.* lateral sizes of 2D PbTe NPLs compared to a sizing curve for spherical PbTe NCs by Peters *et al.*[1] 2D PbTe NPLs exhibit their first excitonic absorbance at higher energies compared to spherical PbTe NCs of the same diameter, emphasizing the additional vertical quantum confinement in the NPLs. It should be noted, however, that the determined values for the energy of the first excitonic transition in 2D PbTe NPLs are error-prone due to rather weakly expressed absorbance feature (see Figure 2a of the main manuscript). This is caused by the overlap of the band edge peak transition with different allowed transitions directly following the first exciton.[1,2]

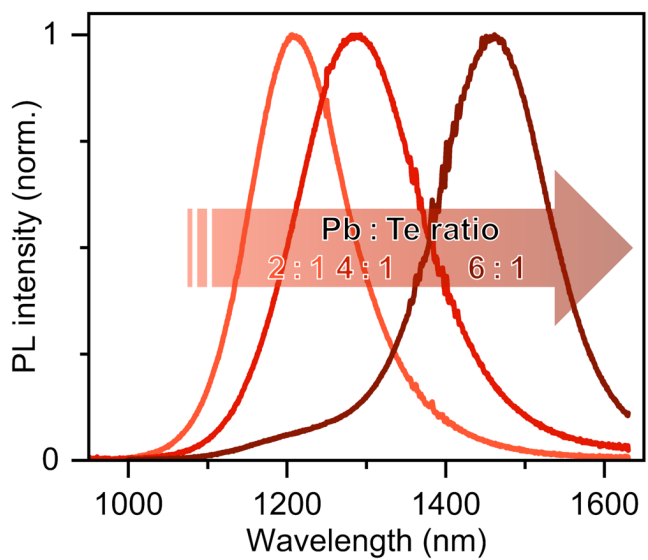

**Figure S5.** PL spectra of PbTe NPLs synthesized with three different Pb:Te ratios of 2:1, 4:1, and 6:1. The PL position is synthetically tunable by changing the Pb:Te ratio, with a higher amount of lead precursor yielding thicker and larger NPLs. TEM images and corresponding size histograms are shown in Figure 1 of the main manuscript as well as in Figure S1, resp.

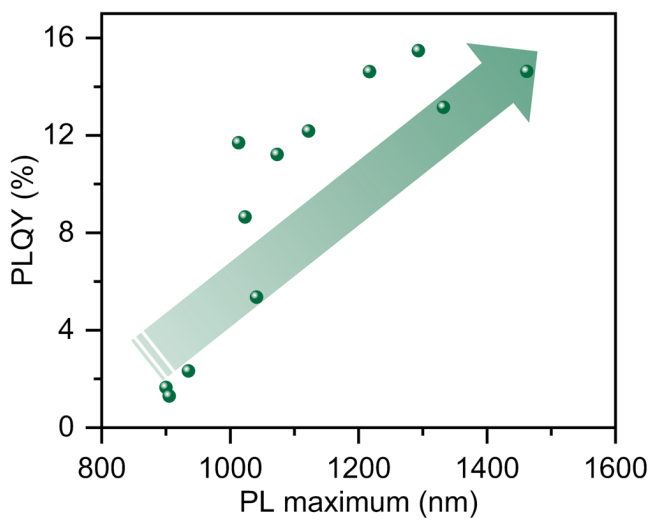

**Figure S6.** PLQY of PbTe NPLs *vs.* the associated PL maxima. The PLQY gradually increases from 1 % at 910 nm (1.36 eV) to 15 % for PL above 1240 nm (1.0 eV).

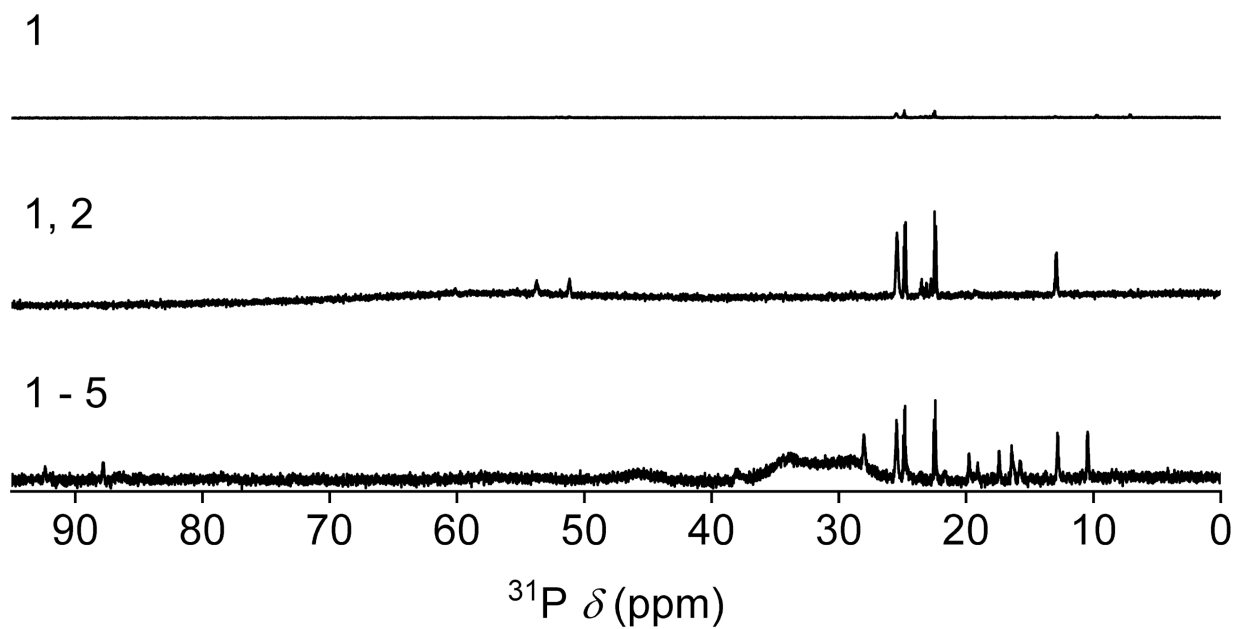

**Figure S7.** $^{31}P\{^1H\}$ NMR spectrum of pure tris(dimethylamino)phosphine (P(N(CH$_3$)$_2$)$_3$ (1)), tris(dimethylamino)phosphine telluride (TeP(N(CH$_3$)$_2$)$_3$ / P(N(CH$_3$)$_2$)$_3$ (2)), and transaminated tris(dimethylamino)phosphine telluride (TeP(N(CH$_3$)$_2$)$_{3-x}$(NHR)$_x$ / P(N(CH$_3$)$_2$)$_{3-x}$(NHR)$_x$, R = C$_8$H$_{17}$ (3 - 5)) in CDCl$_3$ in the range of 0 – 95 ppm corresponding to the $^{31}$P NMR spectra shown in Figure 3 of the main manuscript.

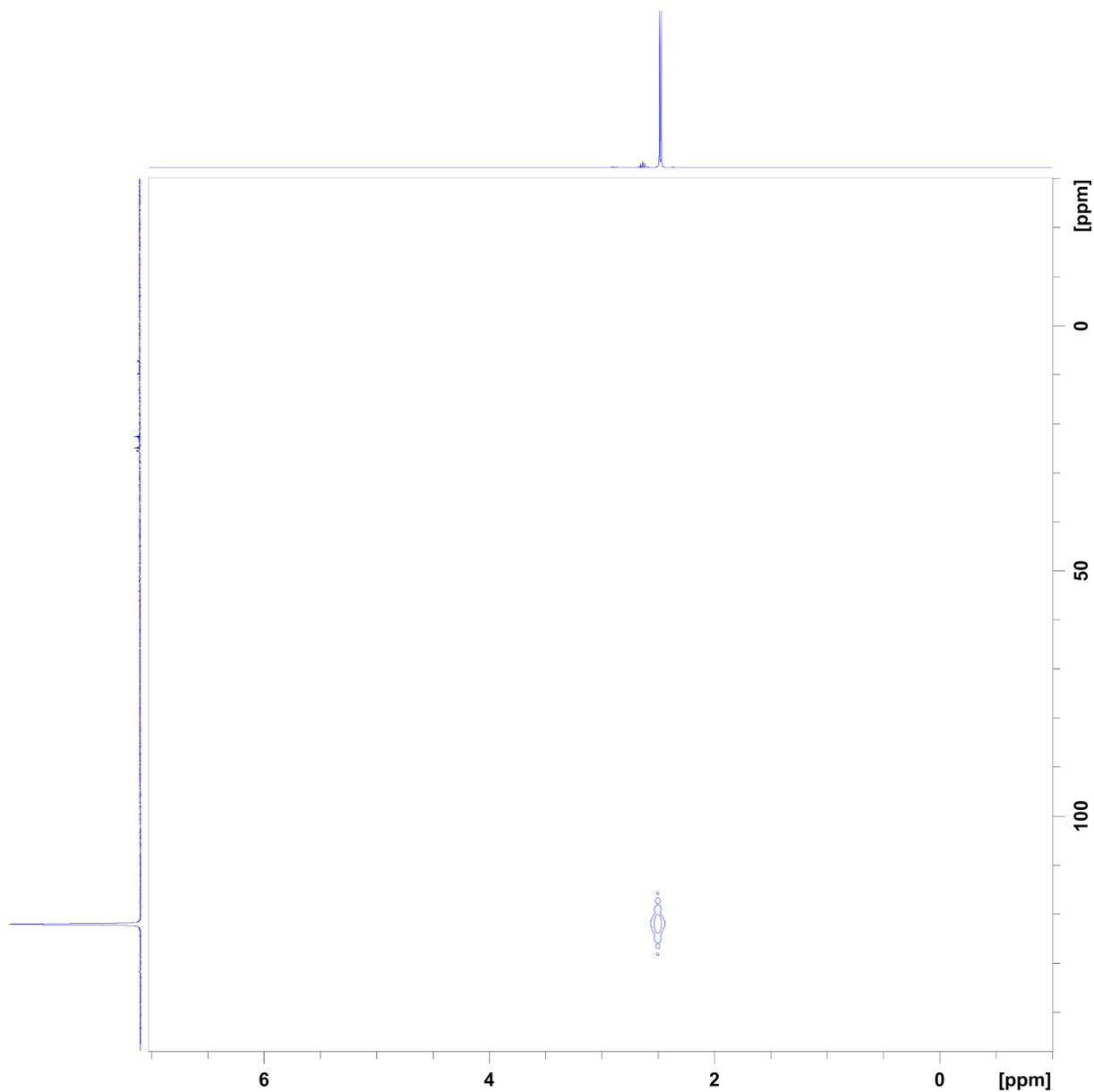

**Figure S8.** $^1$H-$^{31}$P HMBC NMR spectrum of pure tris(dimethylamino)phosphine (P(N(CH$_3$)$_2$)$_3$ (1)) in CDCl$_3$, showing that the $^1$H doublet resonance at 2.48 ppm is linked to the single sharp 122 ppm resonance in the $^{31}$P NMR.

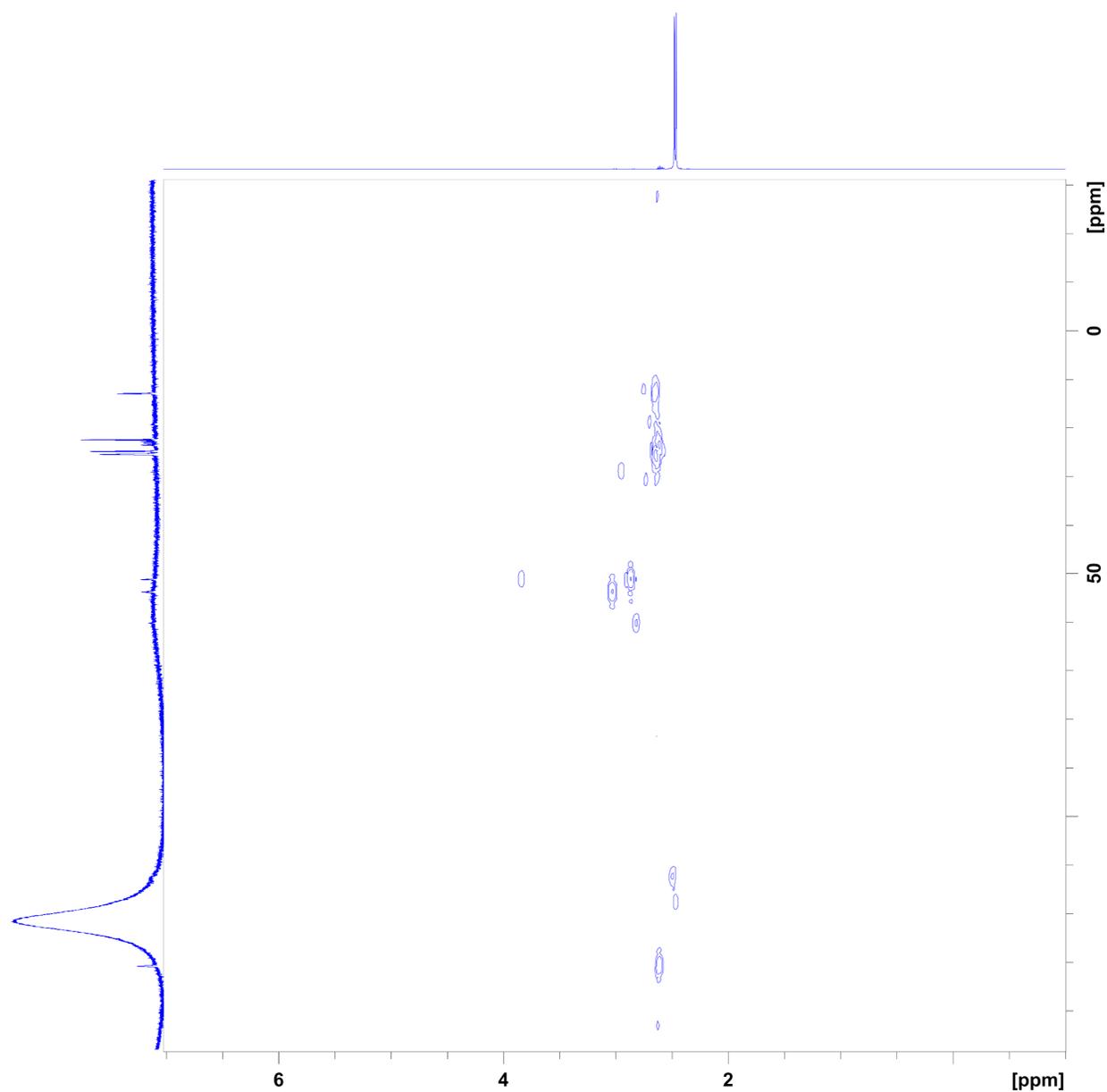

**Figure S9.** $^1$H-$^{31}$P HMBC NMR spectrum of tris(dimethylamino)phosphine telluride (TeP(N(CH$_3$)$_2$)$_3$ / P(N(CH$_3$)$_2$)$_3$ (2)) in CDCl$_3$, showing that the $^1$H doublet resonance at 2.47 ppm is related to the broadened 122 ppm resonance in the $^{31}$P NMR.

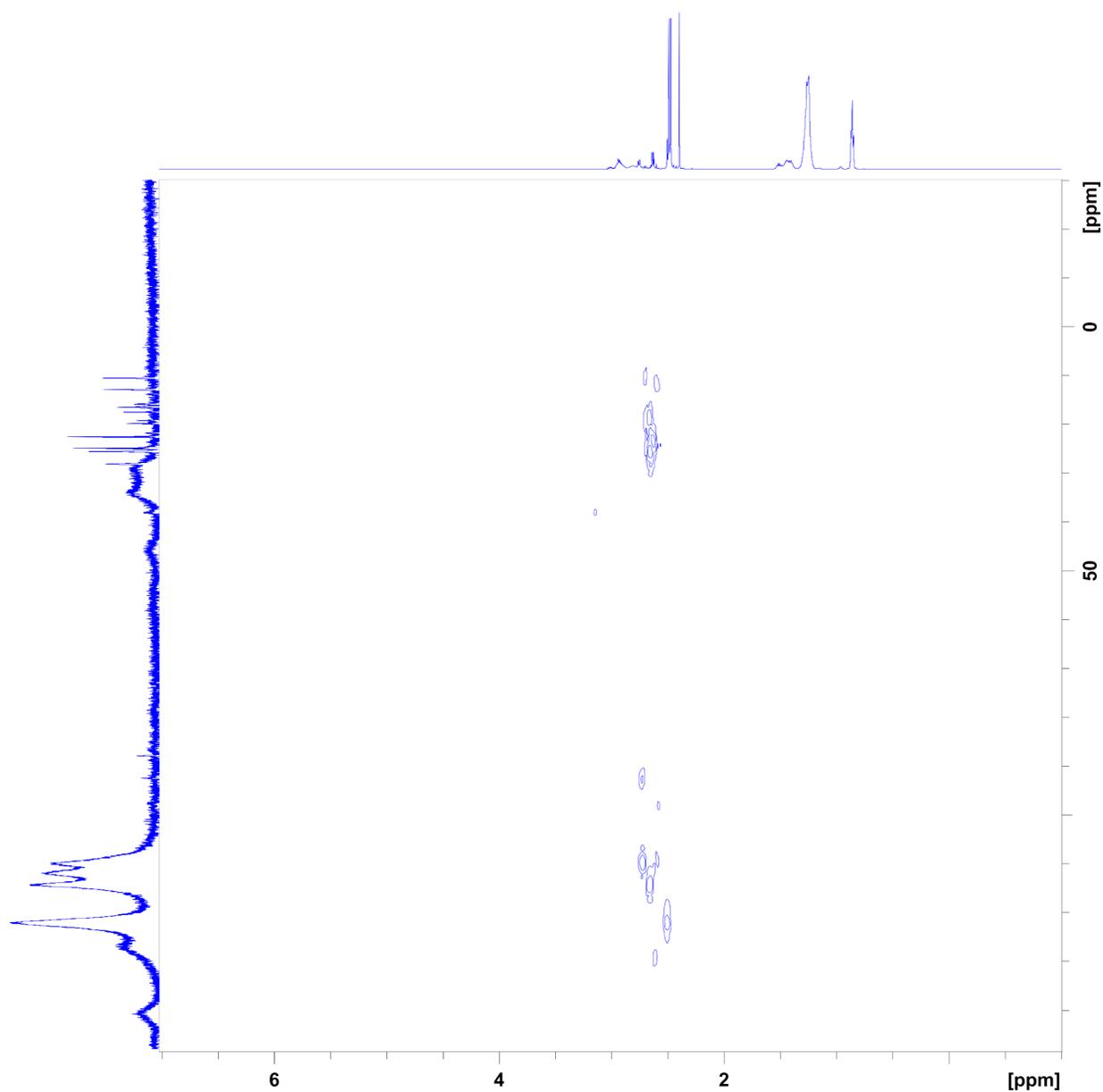

**Figure S10.** $^1$H-$^{31}$P HMBC NMR spectrum of transaminated tris(dimethylamino)phosphine telluride (TeP(N(CH$_3$)$_2$)$_{3-x}$(NHR)$_x$ / P(N(CH$_3$)$_2$)$_{3-x}$(NHR)$_x$, $R$ = C$_8$H$_{17}$ (3 - 5)) in CDCl$_3$, showing the link between the $^1$H doublet resonance at 2.48 ppm and the three $^{31}$P resonances at 110 ppm, 112 ppm, and 114 ppm.

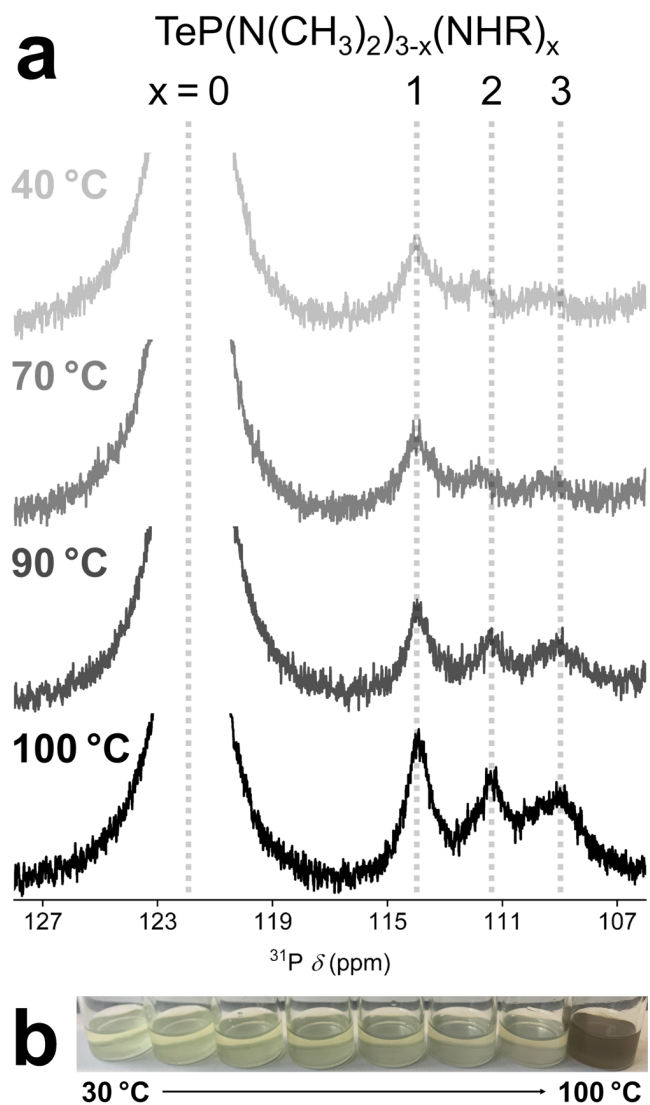

**Figure S 11.** (a) $^{31}$P{$^{1}$H} NMR spectra of aliquots from a tris(dimethylamino)phosphine telluride transamination reaction mixture at different temperatures (40 °C, 70 °C, 90 °C, and 100 °C). The three $^{31}$P resonances corresponding to the single, double, and triple transaminated aminophosphine arise in order when heating up from 40 °C to 100 °C. The spectra show that partial transamination occurs already at 40 °C (and 70 °C), however especially the double and triple transaminated phosphine telluride are largely formed at elevated temperatures above 90 °C, emphasizing the necessity of *ex situ* transamination. (b) Images of collected aliquots from the transamination

reaction, showing a distinct color change upon reaching 100 °C, further underpinning the need for elevated temperatures.

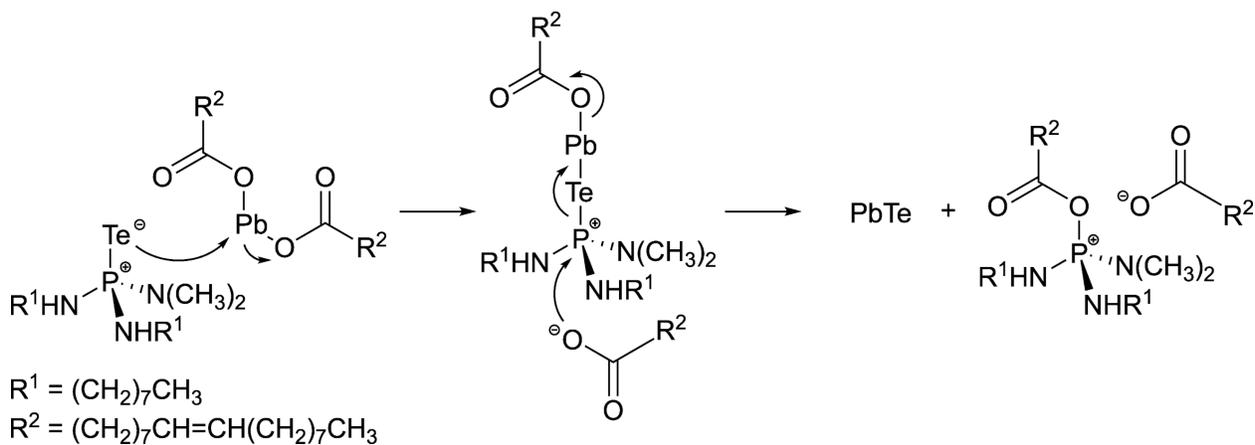

$R^1$ = $(CH_2)_7CH_3$
$R^2$ = $(CH_2)_7CH=CH(CH_2)_7CH_3$

**Figure S12**. Nucleophilic substitution reaction pathway for the formation of PbTe NPLs from lead oleate and transaminated aminophosphine telluride in analogy to the proposed reaction mechanism by Sun *et al*.[3] for the formation of CdTe.

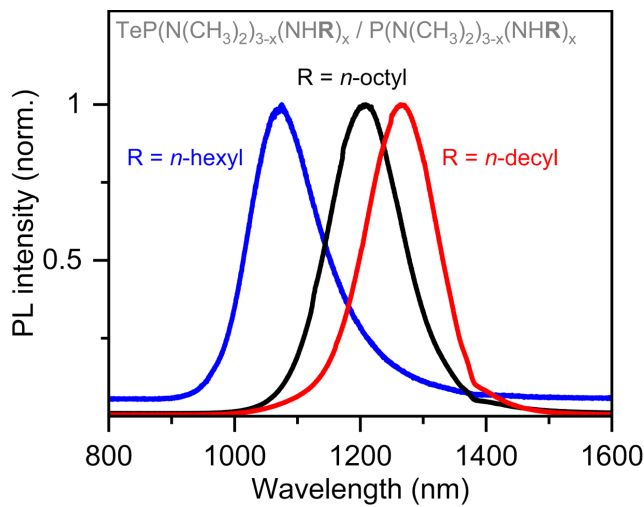

**Figure S13.** PL spectra of PbTe NPLs synthesized using aminophosphine telluride precursors transaminated with three different linear aliphatic primary amines (*n*-hexylamine, *n*-octylamine, *n*-decylamine). Longer alkyl chains reduce the reactivity of the precursor, presumably due to steric hindrance, which results in the formation of thicker and larger NPLs with PL further in the NIR.

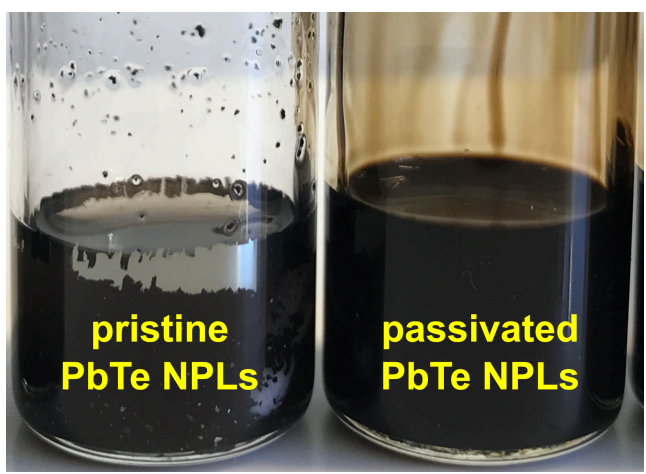

**Figure S14.** Image of pristine and PbI$_2$ passivated PbTe NPLs 24 h after synthesis. While pristine PbTe NPLs change their color from dark brown to black and metallic/elemental

tellurium forms at the inside of the vial, PbI$_2$ passivated NPLs are stable and their dark brown color, colloidal stability and optical properties are retained.